\documentclass[%
 reprint,
 amsmath,amssymb,
 aps,
]{revtex4-2}

\usepackage{graphicx}
\usepackage{dcolumn}
\usepackage{bm}

\newcommand{\chandra}{\textit{Chandra}}
\newcommand{\nustar}{\textit{NuSTAR}}

\newcommand{\ztf}{ZTF~J1901+1458}

\newcommand{\apjs}{Astrophys. J., Suppl. Ser.}

\newcommand{\fe}{\ensuremath{{}^{57}\textrm{Fe}}}

\begin{document}

\title{Constraining Axions with ZTF~J1901+1458}

\author{Leesa Fleury}%
 \email{lfleury@phas.ubc.ca}
\affiliation{%
 Department of Physics and Astronomy, University of British Columbia, Vancouver, BC V6T 1Z1, Canada
}%
\author{Ilaria Caiazzo}%
 \email{ilariac@caltech.edu}
\affiliation{%
 TAPIR, Walter Burke Institute for Theoretical Physics, Mail Code 350-17, Caltech, Pasadena, CA 91125, USA
}%
\author{Jeremy Heyl}%
 \email{heyl@phas.ubc.ca}
\affiliation{%
 Department of Physics and Astronomy, University of British Columbia, Vancouver, BC V6T 1Z1, Canada
}%

\date{\today}


\begin{abstract}
The axion-nucleon coupling enables the production of axions through the decay of excited \fe\ isotopes, and axions produced in the Sun through this process are often a target of helioscope searches. 
We show for the first time that hot, highly magnetic white dwarfs such as \ztf\ are a viable target to search for the X-ray signature of axions that were produced by the \fe\ transition in the core and then converted to photons in the magnetosphere.
We calculate that a 100 ks observation of \ztf\ with \nustar\ would constrain the coupling of axions to nucleons and photons at a level below the bounds of both current and future planned helioscopes.
\end{abstract}

\maketitle


\section{\label{sec:intro}Introduction}

The X-ray emission from magnetic hot white dwarfs may reveal evidence for axions or axion-like particles \cite{2019PhRvL.123f1104D,2021arXiv210412772D}, which have been a major focus of studies to go beyond the Standard Model and to explain dark matter. The QCD axion, proposed to solve the strong CP problem \cite{Peccei:1977hh,Peccei:1977ur,Weinberg:1977ma,Wilczek:1977pj}, is a well motivated addition to the Standard Model.
Axion-like particles, which are pseudo-scalar particles with properties similar to the QCD axion but that do not necessarily relate to the strong CP problem,
also arise naturally in many other extensions to the Standard Model, such as compactified string theories \cite{Witten:1984dg,Conlon:2006tq,Arvanitaki:2009fg,Acharya:2010zx,Higaki:2011me,Cicoli:2012sz,Demirtas:2018akl,Mehta:2021pwf}.

Interactions of axions with photons and nucleons are generic features of QCD axion models, including the benchmark KSVZ \citep{Kim:1979if,Shifman:1979if} and DFSZ \citep{Dine:1981rt,Zhitnitsky:1980tq} models, and are common features of axion-like particle models (see e.g. \citep{DiLuzio:2020wdo} for a recent review).
Many axion models also include a coupling of axions to electrons, including the DFSZ model. 

Axions couple to photons through the interaction term $\mathcal{L} \supset - g_{a\gamma\gamma} a F \tilde{F}$, where $g_{a\gamma\gamma}$ is the axion-photon coupling constant, $a$ is the axion field, and $F$ is the electromagnetic field strength tensor. 
The axion interaction with a fermion species occurs through the operator $\mathcal{L} \supset - i g_{aff} a \bar{f} \gamma_5 f$, where $g_{aff}$ is the axion-fermion coupling constant and $f$ is the fermion field, such as for electrons, protons, or neutrons (i.e. $f=e,p,n$). 
An effective coupling constant $g_{aNN}^\mathrm{eff}$ can also be defined for the axion interaction with the nucleon doublet field $N=(p,n)^T$ in terms of the axion-proton and axion-neutron couplings. 
We define this coupling as $g_{aNN}^\mathrm{eff} = 0.16 g_{app} + 1.16 g_{ann}$ following \cite{2022EPJC...82..120D}.

The axion couplings to photons and fermions each have a model-dependent relation to the axion mass, $m_a$.
To summarize, typical values for the QCD axion couplings are \citep{2007JPhA...40.6607R}
\begin{eqnarray*}
    g_{aNN}^{\textrm{\scriptsize eff}} &=& 1.6 \times 10^{-8} \left (\frac{m_a}{1\textrm{eV}} \right ) ~~\textrm{(KSVZ)} 
    \label{eq:gaNNKSVZ} \\
    g_{a\gamma\gamma} &=& 4 \times 10^{-10} \left ( \frac{m_a}{1\textrm{eV}} \right ) \textrm{GeV}^{-1} ~\textrm{(KSVZ)}
    \label{eq:gaggKSVZ} \\
    g_{aee} &\approx& 3 \times 10^{-11} \left ( \frac{m_a}{1\textrm{eV}} \right )~~\textrm{(DFSZ)},
    \label{eq:gaggDFSZ}
\end{eqnarray*}
but axion-like particles can have couplings much larger than these \citep{DiLuzio:2020wdo}.
The axion-nucleon coupling for the DFSZ model of the QCD axion can also be up to a factor of $\sim 4$ larger than for the KSVZ model \cite{2022EPJC...82..120D}.
For both benchmark QCD axion models, the axion-nucleon coupling is typically three orders of magnitude larger than the DFSZ axion-electron coupling.
As X-ray observations of white dwarfs searching for evidence of axions have focused on axions produced through the axion-electron interaction, the greater strength of the axion-nucleon interaction highlights the potential power of similar X-ray searches for evidence of axions produced through nuclear interactions.

Astrophysical observations are often used to search for signatures indicative of the various possible axion interactions with Standard Model particles and to constrain the axion coupling constants. 
The axion-nucleon coupling has been probed indirectly through the observed neutrino emission from SN 1987A \cite{Turner:1987by,Burrows:1988ah,Raffelt:1987yt,Raffelt:1990yz,Carenza:2019pxu,Carenza:2020cis,Fischer:2021jfm} and the cooling of neutron stars \cite{Keller:2012yr,Sedrakian:2015krq,Hamaguchi:2018oqw,Beznogov:2018fda,Sedrakian:2018kdm,Leinson:2014ioa}.
Helioscope experiments, which search for axions produced in the Sun, have also been used to constrain the axion-nucleon coupling. 
Axions can be produced in the Sun through the decay of excited nuclear states such as the first excited state of \fe. 
The CERN Axion Solar Telescope (CAST \citep{Arik:2013nya,Arik:2015cjv}) has searched for axions produced in this way, setting current constraints for the axion-nucleon coupling \cite{CAST:2009jdc,2022EPJC...82..120D}.
A similar search has been proposed for the future planned International Axion Observatory (IAXO \citep{2022EPJC...82..120D,BabyIAXO:2020mzw,IAXO:2019mpb}), along with improved calculations for the axion flux from the \fe\ transition \cite{2022EPJC...82..120D} using updated nuclear matrix elements \cite{Avignone:2017ylv}.

White dwarfs are another popular target of searches for axions produced from astrophysical sources and have typically been used to probe the axion-electron coupling, which enables the production of axions within a white dwarf through axion bremsstrahlung.
This extra source of energy loss would modify the cooling of white dwarfs and thus the white dwarf luminosity function, which in comparison to the observed luminosity function has been used to constrain the axion-electron coupling \cite{2015ApJ...809..141H,2016ApJ...821...27G,Isern:2008nt,Isern:2008fs,Bertolami:2014wua}. 
Furthermore, in the strong field surrounding magnetic white dwarfs, axions can convert to photons and vice versa. 
Axions produced by bremsstrahlung which then convert to photons in the surrounding magnetic field produce a black-body-like spectrum in the X-ray, which was the focus of a 100-ks observation of the magnetic white dwarf RE~J0317-853 \citep{2021arXiv210412772D} with \chandra.

In this work, we show that hot, highly magnetized white dwarfs are also ideal targets to probe the axion-nucleon coupling via the \fe\ transition, and can in fact give bounds below those of both current and planned future helioscope constraints of axions produced in the Sun through nuclear transitions.
If the temperature of the white dwarf is sufficiently high, low-lying nuclear states may be excited within the star, and these states may decay through the emission of an axion.  
Among all of the low-lying nuclear states, the excited state of \fe\ at 14.4~keV stands out through the combination of low energy and relatively large abundance within the star \cite{SolarElementalAbundances}.  
In general, these axions would stream out of the star unimpeded and unnoticed. However, in the case of a strongly magnetized white dwarf like \ztf\ \cite{2021Natur.595...39C}, these axions have a chance to transform into X-ray photons at 14.4~keV as they pass through the magnetosphere of the white dwarf.  
White dwarfs are ideal targets for measurements of this signature because their thermal emission in the hard X-rays is negligible, and therefore they provide the chance for a very clean detection.
We demonstrate the potential constraining power of X-ray searches for the \fe\ axion signal from magnetic white dwarfs by calculating the projected constraints for the axion-nucleon and axion-photon couplings that could be obtained with a 100~ks observation of \ztf\ by \nustar.

\section{\label{sec:calculations}Calculations}

The spectrum of axions produced through the \fe\ nuclear transition is a narrow peak at the nuclear excitation energy, $E^* = 14.4$~keV. 
The resultant photon number flux at Earth induced by axions being produced via the nuclear transition process in the core of a white dwarf and then converting to photons in the magnetosphere is given by \cite{2019PhRvL.123f1104D}
\begin{equation} \label{eq:flux}
    \Phi_{\gamma} = \mathcal{N}_a \ M_\mathrm{WD} \times p_{a \rightarrow \gamma} \times \frac{1}{4 \pi d_\mathrm{WD}^2} 
    \quad ,
\end{equation}
where $\mathcal{N}_a$ is the emission rate per unit mass of axions produced in the white dwarf interior (which we consider to be isothermal),
$M_\mathrm{WD}$ is the white dwarf mass, $p_{a \rightarrow \gamma}$ is the probability of an axion with energy $E^*$ converting into a photon in the magnetosphere of the white dwarf, and $d_\mathrm{WD}$ is the distance to the white dwarf from the point of observation.
The product $\mathcal{N}_a M_\mathrm{WD}$ is the mass averaged number of axions produced by the white dwarf per unit time, and multiplying this quantity by $E^*$ yields the luminosity of axion production under the assumption of an isothermal white dwarf core.

Modeling the core of the white dwarf as isothermal, the number density of axions produced from the \fe\ nuclear transition in the white dwarf core is \cite{2022EPJC...82..120D}
\begin{equation} \label{eq:La}
    \mathcal{N}_a = \mathcal{N} \ \omega_1(T_c) \ \frac{1}{\tau_0} \frac{1}{\left(1+\alpha\right)} \  \frac{\Gamma_a}{\Gamma_\gamma}
    \quad ,
\end{equation}
where $\mathcal{N}$ is the \fe\ number density per unit mass of stellar matter, $\omega_1$ is the occupation number of the first excited state at the temperature $T_c$ of the white dwarf core, $\tau_0$ is the lifetime of the excited state, $\alpha$ is the internal conversion coefficient, and $\Gamma_a / \Gamma_\gamma$ is the branching ratio of axion emission relative to photon emission.

The axion emission rate depends on the axion-nucleon coupling through the term $\Gamma_a / \Gamma_\gamma$.
For axions produced by the \fe\ transition, the axion flux from the Sun has recently been calculated by \cite{2022EPJC...82..120D} using the updated nuclear matrix elements of \cite{Avignone:2017ylv}. 
This yielded an updated axion-to-photon branching ratio for ultrarelativistic axions of
\begin{equation}
    \frac{\Gamma_a}{\Gamma_\gamma} = 2.32 \ \left( g_{aNN}^\mathrm{eff} \right)^2 \quad ,
\end{equation}
where $g_{aNN}^\mathrm{eff}$ is the effective axion-nucleon coupling constant, defined as $g_{aNN}^\mathrm{eff} = 0.16 g_{app} + 1.16 g_{ann}$ in terms of the axion couplings to protons and neutrons \cite{2022EPJC...82..120D}.

\begin{figure}
    \centering
    \includegraphics[width=0.9\columnwidth,trim=0 0 0 2cm]{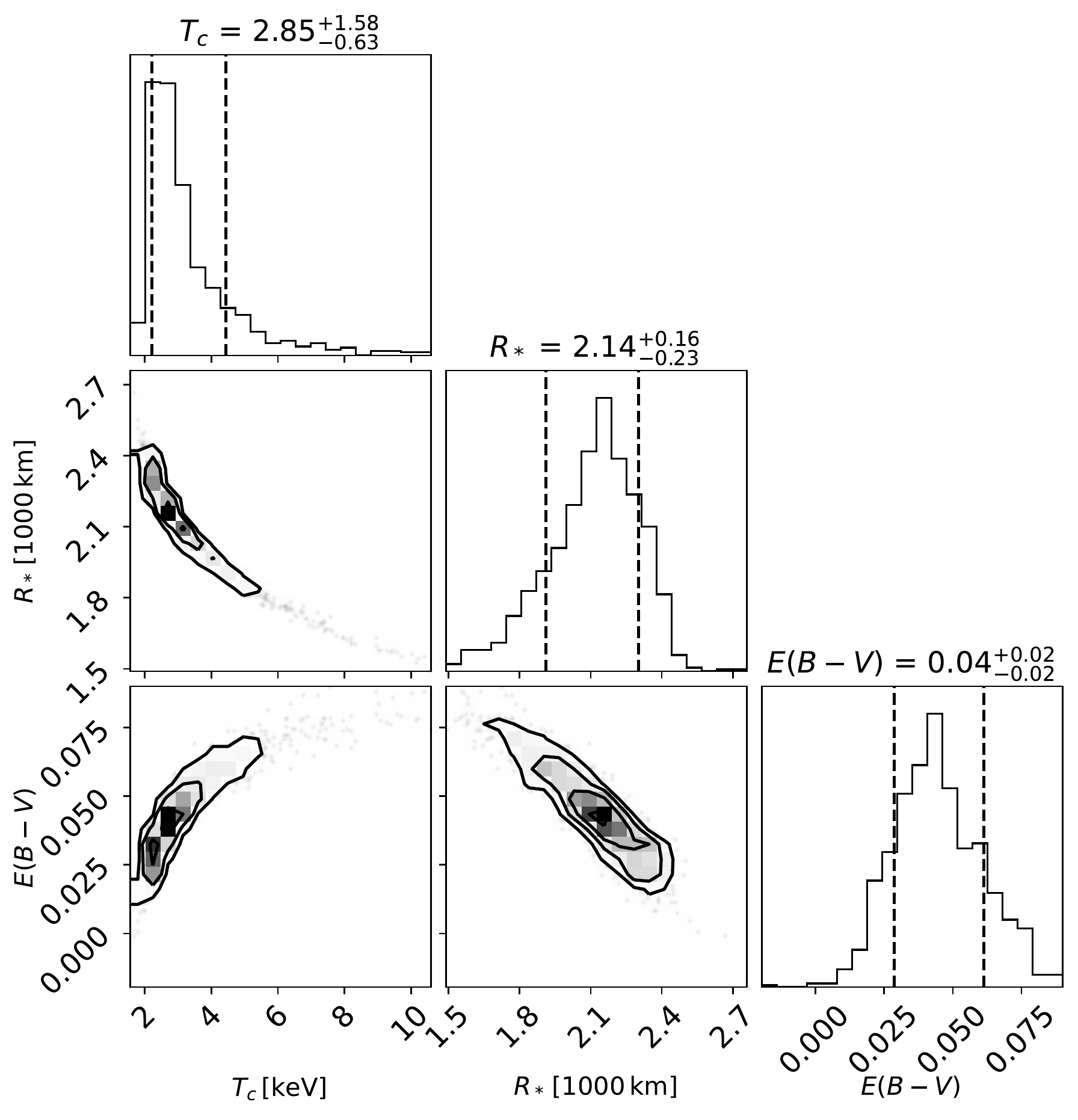}
    \caption{Constraints on the core temperature of \ztf\ from the published photometry \cite{2021Natur.595...39C}.}
    \label{fig:core_temp}
\end{figure}

The core temperature of the white dwarf is another important parameter for the calculation of $\mathcal{N}_a$, as the occupation number $\omega_1$ is temperature-dependent.
We determined the core temperature of \ztf\ using the published photometry and fitting technique of \cite{2021Natur.595...39C}, with $T_c$ used as one of the free parameters instead of the effective temperature and implementing the relation between the core temperature and photon luminosity \cite{2019PhRvL.123f1104D}
\begin{equation}
    k T_c \simeq \left( 0.3 \ \mathrm{keV} \right) \left( \frac{L_\gamma}{10^{-4} \ L_\odot} \right)^{0.4}
    \quad ,
\end{equation}
where $k$ is the Boltzmann constant.
The results of the joint fit for $T_c$ along with the radius of the white dwarf, $R_*$, and colour excess due to interstellar reddening, $E(B-V)$, are shown in Fig.~\ref{fig:core_temp}.
Based on these results, we use a core temperature of 
$k T_c = 2.85^{+1.58}_{-0.63}~\textrm{keV}$ 
for our calculation of the axion emission rate.

The thermal occupation number of an excited state with excitation energy $E^*$ as a function of temperature $T$ is 
$\omega_1 = (2 J_1 + 1) e^{-E^*/kT} \ / \ [(2 J_0 + 1) + (2 J_1 + 1) e^{-E^*/kT}]$, 
where $J_0$ and $J_1$ are the angular momenta of the ground and excited state, respectively \cite{2022EPJC...82..120D}. 
These angular momenta are $J_0=1/2$ and $J_1 = 3/2$ for the \fe\ ground and first excited state, respectively \cite{Roehlsberger:2004}, giving a thermal occupation number of $\omega_1 = 2 e^{-E^*/kT} / \left(1 + 2 e^{-E^*/kT}\right)$. 
This further simplifies to approximately $\omega_1 \sim 2 e^{-E^*/kT_c}$ for the core temperature of \ztf\ and an isothermal core.

A large core temperature can also broaden the width of the expected axion signal.
In the case of the Sun, the energy of the axion is broadened by the thermal motion of the nuclei, yielding a width of a few eV \cite{2022EPJC...82..120D}. 
As the tempreature in the core of the white dwarf is larger than in the Sun, so is the thermal broadening (5~eV for ZTF~J1901+1458); however, the varying gravitational potential through the core has a larger effect through the gravitational redshift that is about 280~km/s at the surface and 980~km/s at the centre, yielding a width of 33~eV, which dominates over the thermal effects.
These broadening effects are negligible compared to the spectral resolution of \nustar\ ($\approx 400$~eV), so we work in terms of the total flux rather than the spectral flux and approximate that all of the axions are emitted with the same energy $E^*$.

The values for all of the other parameters used in the calculation of $\mathcal{N}_a$ are taken from the literature.
For the parameters characterizing the first excited state of \fe, we use values from \cite{Roehlsberger:2004} of $\tau_0 = 141~\textrm{ns}$ and $\alpha = 8.56$.
To get the number density of \fe\ nuclei, we use the abundances reported by \cite{SolarElementalAbundances}. 
We use a proto-solar hydrogen mass fraction of 0.71 and
a number fraction of \fe\ nuclei relative to protons of $7.34 \times 10^{-7}$ (see Table 9 of \cite{SolarElementalAbundances}). 
The corresponding \fe\ number density per unit mass is $\mathcal{N} = 6.24 \times 10^{50}~M_\odot^{-1}$.
Furthermore, we performed stellar evolution simulations using Modules for Experiments in Stellar Astrophysics (MESA \cite{Paxton2011,Paxton2013,Paxton2015,Paxton2018,Paxton2019}) to verify that stellar evolution processes do not alter the iron abundance in the stellar core.
Finally, for the white dwarf parameters needed in the rest of the flux calculation, we use the values for \ztf\ reported in \cite{2021Natur.595...39C}, for which the mass is $M_\mathrm{WD} = 1.346\pm0.019~M_\odot$ and the distance is $d_\mathrm{WD} = 41.4\pm0.1$~pc.

To determine the observable photon flux $\Phi_{\gamma}$ that would be induced by axions produced at the rate $\mathcal{N}_a$ in the white dwarf, we must also calculate the probability of the axions converting into photons as they propagate outward through the magnetic field surrounding the white dwarf.
Under the approximation that the axions travel along radial trajectories (relative to the center of the white dwarf), the propagation of the axion-photon field is described by the equations \cite{Raffelt:1987im}
\begin{equation}
    \left[ i \partial_r + E +
    \begin{pmatrix}
    \Delta_\parallel & \Delta_B\\
    \Delta_B & \Delta_a
    \end{pmatrix}
    \right]
    \begin{pmatrix} A_\parallel \\ a \end{pmatrix}
    = 0
    \quad ,
    \label{eq:axion-photon}
\end{equation}
\begin{align*}
    &\Delta_{\parallel}(r) = \left(7/2\right) \left(\alpha_\mathrm{EM} / 45\pi \right) E \left[ B(r) / B_\mathrm{crit} \right]^2 \sin^2\Theta \\
    &\Delta_B(r) = (1/2) \ g_{a\gamma\gamma} \ B(r) \sin\Theta \\
    &\Delta_a = - m_a / (2 E)
    \quad ,
\end{align*}
where $r$ is the radial component, $E \approx E^*$ is the axion energy, $a(r)$ is the axion field, 
$A_{\parallel}(r)$ is the component of the electromagnetic vector potential that is parallel to the external magnetic field (in the plane whose normal vector is aligned with the direction of propagation),
$\alpha_\mathrm{EM} = 1 / 137$ is the electromagnetic fine structure constant,
$B_\mathrm{crit} = 4.414 \times 10^{13}$~G is the quantum electrodynamics critical field strength,
$\Theta$ is the angle between the magnetic field and the radial propagation direction, and
$B(r)$ is the magnetic field strength at radius $r$.
For details on the origin and solution of the axion-photon propagation equations for magnetic stars, see e.g. \cite{Raffelt:1987im,2006PhRvD..74l3003L}.

To calculate the probability of an axion converting to a photon in the external white dwarf magnetic field, we solve Eq.~\ref{eq:axion-photon} numerically for an initial pure axion state. 
The probability $p_{a \rightarrow \gamma}$ is given by the squared magnitude of the asymptotic solution for $A_\parallel$.
For this calculation, we model the external magnetic field of the white dwarf as a magnetic dipole, $B(r) = B_0 \left(R_\mathrm{WD} / r \right)^3$, where $B_0$ is the magnetic field at the surface of the white dwarf and $R_\mathrm{WD}$ is the white dwarf radius. 
For this form of the magnetic field, $\sin\Theta$ has a fixed value along the radial trajectory which we take to be unity.
The values of the relevant magnetic field parameters for \ztf\ are $B_0 \sim 800$~MG and $R_\mathrm{WD} = 2,140^{+160}_{-230}$~km \cite{2021Natur.595...39C}.

\begin{table}
    \centering
    \begin{tabular}{l | l}
        Parameter \ & \ Value \\
        \hline
        $E^*$ & \ 14.4~keV \\
        $J_0$ & \ 1/2 \\
        $J_1$ & \ 3/2 \\
        $\tau_0$ & \ 141~ns \\
        $\alpha$ & \ 8.56 \\ 
        $\mathcal{N}$ & \ $6.24 \times 10^{50}~M_\odot^{-1}$ \\ 
        $M_\mathrm{WD}$ & \ 1.34~$M_\odot$ \\ 
        $d_\mathrm{WD}$ & \ 41.44~pc \\ 
        $B_0$ & \ 800~MG \\
        $R_\mathrm{WD}$ & \ 2,100~km \\
        $k T_c$ & \ $2.85^{+1.58}_{-0.63}$~keV \\
    \end{tabular}
    \caption{Parameter values used to calculate the X-ray flux induced by axions produced in the \fe\ nuclear transition process for \ztf. 
    The properties of the first excited state of \fe\ come from \cite{Roehlsberger:2004}.
    The abundances used to determine $\mathcal{N}$ come from \cite{SolarElementalAbundances}. 
    The \ztf\ white dwarf parameters come from \cite{2021Natur.595...39C}, except for $T_c$ which is calculated in this work.}
    \label{tab:params}
\end{table}

The key parameter values used to calculate the X-ray photon flux at Earth are summarized in Table~\ref{tab:params}.
The observable photon flux induced by the \fe\ transition depends on both the axion-nucleon coupling and the axion-photon coupling.
The emission rate of axions produced by the \fe\ nuclear transition in the white dwarf interior goes as $(g_{aNN}^\mathrm{eff})^2$, while the probability of the axions converting into photons in the external magnetic field goes as $(g_{a\gamma\gamma})^2$. 
Thus, observations of the hard X-ray emission from magnetic white dwarfs can constrain the product of the couplings: $g_{aNN}^\mathrm{eff} g_{a\gamma\gamma}$.
Furthermore, the probability of an axion converting to a photon depends on the axion mass, so the constraints on $g_{aNN}^\mathrm{eff} g_{a\gamma\gamma}$ from X-ray observations will be a function of axion mass, which we evaluate numerically over a grid a axion mass values.

We calculate the constraints that could be obtained for $g_{aNN}^\mathrm{eff} g_{a\gamma\gamma}$ from \nustar\ observations with a total exposure time of 100~ks.
Using XSPEC simulations, we determined the minimum flux for a three-sigma detection of a narrow (width of 0.03~keV), Gaussian spectral line at 14.4~keV above the background (30-arcsecond extraction region and one arcminute off-axis) to be $1.6\times 10^{-6}$~photons~s$^{-1}$~cm$^{-2}$. In the background, we also included the thermal emission from the white dwarf atmosphere modelled as a blackbody at a temperature $k T_\textrm{eff}=3.88$~eV with a radius of 2,100~km.
For a given axion mass, the value of $g_{aNN}^{\textrm{\scriptsize eff}} g_{a\gamma\gamma}$ for which the photon number flux given by Eq.~\ref{eq:flux} equals (or exceeds) the detector threshold sets the constraint.

\section{\label{sec:results}Results}

\begin{figure}
    \centering
    \includegraphics[width=\columnwidth]{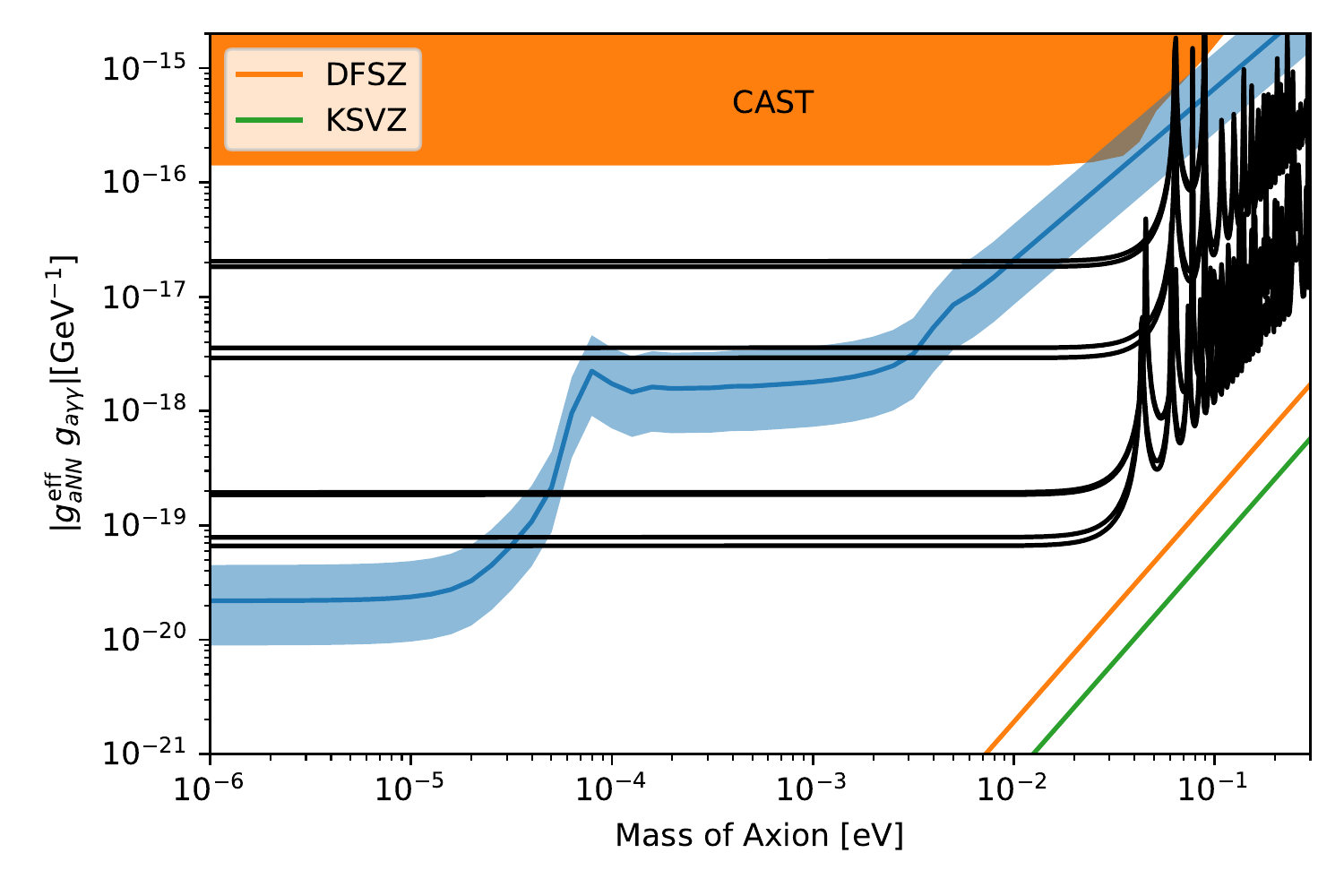}
    \caption{Projected constaints on the product of the axion-nucleon coupling with the axion-photon coupling that would be obtained with a 100~ks NuSTAR observations in blue.  The band accounts for the uncertainty in the core temperature of the white dwarf. 
    The black curves give limits that may be obtained with the future International Axion Observatory \citep{2022EPJC...82..120D,BabyIAXO:2020mzw,IAXO:2019mpb}.  
    The upper four are for the intermediate-stage BabyIAXO, and the lower four are for the fully operational IAXO.}
    \label{fig:constraint}
\end{figure}

The potential \nustar\ signature of axion production through nuclear processes is a narrow emission line at 14.4~keV (the excitation energy of the \fe\ nucleus).  
The projected axion constraints that would be obtained from a 100~ks \nustar\ observation of the white dwarf \ztf\ are depicted in Fig.~\ref{fig:constraint}, where they are compared to constraints from current and future observatories. 
These constraints would be more stringent than the existing constraints from CAST for all axion masses and even more stringent than the ones that will be obtained by the future proposed IAXO by an order of magnitude for small axion masses. 
We show the limits for the four different configurations of both the proposed intermediate-stage BabyIAXO and the proposed fully operational IAXO (see Tab.~I of Ref.~\cite{2022EPJC...82..120D} for a summary of these configurations).

In addition to constraining $g_{aNN}^\mathrm{eff} g_{a\gamma\gamma}$, X-ray observations of \ztf\ could also constrain $g_{aee} g_{a\gamma\gamma}$ as white dwarfs can produce axions through electron bremsstrahlung, which yields a blackbody-like spectrum of axions. 
This was the focus of a 100-ks observation of a cooler magnetic white dwarf, RE~J0317-853 \citep{2021arXiv210412772D}, with \chandra.
These axions are typically produced at lower energies (near the temperature of the core at a few keV), so the resulting X-rays lie squarely in the energy range probed by \chandra, but constraints could be derived from \nustar\ observations as well. 

\ztf\ is a prime target to search for axion signatures in the X-rays because it is one of the hottest and most strongly magnetized white dwarfs known. 
Both properties increase the predicted strength of the \fe\ nuclear transition line and bremsstrahlung emission, and therefore the chance of detection, or the significance of the constraints in case of a non-detection.
Recently published photometry and spectroscopy \citep{2021Natur.595...39C} yield somewhat broad constraints on the core temperature of the white dwarf (see Fig.~\ref{fig:core_temp}) that result in a broad range for the implied constraints on the axion (the shadowed blue area in Fig.~\ref{fig:constraint}).  Fortunately, follow-up ultraviolet spectroscopy observations are scheduled for the current HST cycle \citep{2021hst..prop16753C} that should provide stronger constraints on the effective temperature and therefore the core temperature of the white dwarf, which would narrow the band of constraints that could be achieved with \nustar\ observations.

\section{\label{sec:conclusions}Conclusions}

White dwarfs are a popular target of indirect searches for axions. 
While observations of white dwarfs have been used to constrain the coupling of axions to electrons, white dwarfs have not previously been identified as a target for probing the axion coupling to nucleons. 
In this work, we have shown for the first time that observations of white dwarfs can be used to probe the axion-nulceon coupling through searches for the X-ray signal from hot, highly magnetized white dwarfs that would be induced by the process of axions being produced in the core via the nuclear transition of the first excited state of \fe\ and then converting into photons in the magnetosphere.

The recently discovered white dwarf \ztf\ is a compelling target to search for an X-ray signal arising from axions produced via the \fe\ transition.
We have shown that a 100~ks observation of \ztf\ by \nustar\ would constrain the coupling of axions with nucleons and photons at a level below current constraints for all axion masses and below the constraints that would be obtained by planned future terrestrial experiments at small masses.
This would provide a dramatic improvement in our knowledge of these particles that are critical to our understanding of the Standard Model and possibly dark matter as well.

\begin{acknowledgments}
This work has been supported by the Natural Sciences and Engineering Research Council of Canada through the Discovery Grants program and Compute Canada. I.C. is a Sherman Fairchild Fellow at Caltech and thanks the Burke Institute at Caltech for supporting her research.
\end{acknowledgments}

\bibliography{main}

\end{document}